%% file: main_heep_arxiv.tex
\documentclass[conference, twocolumn]{IEEEtran}
\IEEEoverridecommandlockouts
\def\BibTeX{{\rm B\kern-.05em{\sc i\kern-.025em b}\kern-.08em
		T\kern-.1667em\lower.7ex\hbox{E}\kern-.125emX}}

\usepackage{url}
\input{common/includes_paper.tex}

\input{common/macros_formulae.tex}
\input{common/macros_common.tex}

\makeglossaries

\input{common/glossary_entries.tex}	
\glsdisablehyper

\begin{document}

\title{Implicit Incorporation of Heuristics in MPC-Based Control of a Hydrogen Plant
}

\author{Thomas Schmitt$^{1}$, 
	Jens Engel$^{1}$,
	Martin Kopp$^{2}$, 
	Tobias Rodemann$^{1}$
	\thanks{
		$^{1}$Honda Research Institute Europe GmbH, 
		Offenbach, Germany. 
		E-mail: {\tt\footnotesize \{thomas.schmitt, jens.engel, tobias.rodemann\}@honda-ri.de}
	}
	\thanks{
		$^{2}$Honda R\&D Europe (Deutschland) GmbH, 
		Offenbach, Germany. 
		E-mail: {\tt\footnotesize martin\_kopp@de.hrdeu.com}
	}%
	\thanks{Funded by HA Hessen Agentur GmbH}
}

\maketitle

\begin{abstract}
	The replacement of fossil fuels in combination with an increasing share of renewable energy sources leads to an increased focus on decentralized microgrids. 
	One option is the local production of green hydrogen in combination with \glspl{fcv}. 
	In this paper, we develop a control strategy based on \gls{mpc} for an \gls{ems} of a hydrogen plant, which is currently under installation in Offenbach, Germany. 
	The plant includes an electrolyzer, a compressor, a low pressure storage tank, and six medium pressure storage tanks with complex heuristic physical coupling during the filling and extraction of hydrogen. 
	Since these heuristics are too complex to be incorporated into the \gls{ocp} explicitly, we propose a novel approach to do so implicitly. 
	First, the \gls{mpc} is executed without considering them. 
	Then, the so-called allocator uses a heuristic model (of arbitrary complexity) to verify whether the \gls{mpc}'s plan is valid. 
	If not, it introduces additional constraints to the \gls{mpc}'s \gls{ocp} to implicitly respect the tanks' pressure levels.
	The \gls{mpc} is executed again and the new plan is applied to the plant.
	Simulation results with real-world measurement data of the facility's energy management and realistic fueling scenarios show its advantages over rule-based control. 
\end{abstract}


\glsresetall
\input{inc/01_intro.tex}
\input{inc/02_system_overview.tex}
\input{inc/03_control_approach.tex}
\input{inc/04_simulation_results.tex}
\input{inc/05_conclusion.tex}

\FloatBarrier
\newcommand{\BIBdecl}{\setlength{\itemsep}{-6pt}}
\bibliographystyle{IEEEtran}
\bibliography{common/lib_heep}
	
\end{document}

%% file: common/includes_paper.tex
\usepackage{amsmath,amssymb,amsfonts}
\usepackage[ruled,norelsize]{algorithm2e}
\usepackage{graphicx}
\usepackage{textcomp}
\usepackage{xcolor}

\usepackage[english]{babel}
\usepackage[utf8]{inputenc}
\usepackage[font=small]{caption}

\usepackage{placeins}
\usepackage{color}
\usepackage{mathtools}
\usepackage{parskip} 
\usepackage{xspace}
\usepackage{microtype}
\usepackage{cite} 
\usepackage{tabularx}
\usepackage{multirow}

\usepackage{gensymb} 
\usepackage{booktabs}
\usepackage{ifthen}
\usepackage{eurosym} 
\usepackage{amsbsy} 
\usepackage[mathcal]{euscript}

\usepackage{subfigure}

\usepackage{grffile}

\usepackage{icomma}
\usepackage{tabularx}
\usepackage{nicefrac} 
\usepackage{listings} 
	\lstdefinestyle{myCustomMatlabStyle}{
		tabsize=4,
		showspaces=false,
		showstringspaces=false
	}
\usepackage{transparent} 
\usepackage{import} 

\usepackage[acronym]{glossaries} 
\definecolor{myblue}{RGB}{62, 176, 247}
\usepackage[hidelinks]{hyperref}

%% file: common/macros_formulae.tex
\newcommand{\nk}{\ensuremath{(n|k)}\xspace}
\let\k\relax 
\newcommand{\k}{\ensuremath{(k)}\xspace}
\newcommand{\kpo}{\ensuremath{(k+1)}\xspace}

\newcommand{\Np}{\ensuremath{N_\g{p}}\xspace}
\newcommand{\Npred}{\ensuremath{N_\g{pred}}\xspace}

\newcommand{\sumnNp}{\sum_{n=0}^{\Np-1}}
\newcommand{\sumnNppo}{\sum_{n=0}^{\Np}}

\newcommand{\xvek}{\ensuremath{x}\xspace}

\newcommand{\xk}{\ensuremath{\xvek(k)}\xspace}

\newcommand{\xkp}{\ensuremath{\xvek(k+1)}\xspace}

\newcommand{\xseq}{\ensuremath{\boldsymbol{\xvek}}\xspace}

\newcommand{\uvek}{\ensuremath{u}\xspace}

\newcommand{\uk}{\ensuremath{\uvek(k)}\xspace}

\newcommand{\ukk}[1][]{
	\ifthenelse{ \equal{#1}{} }
	{\ensuremath{\uvek(k+1)}\xspace}
	{\ensuremath{\uvek(k+{#1})}\xspace}
} 

\newcommand{\uaux}{\ensuremath{\uvek_\g{aux}}\xspace}
\newcommand{\uauxseq}{\ensuremath{\boldsymbol{\uvek}_\g{aux}}\xspace}

\newcommand{\useq}{\ensuremath{\boldsymbol{\uvek}}\xspace}

\newcommand{\cgridbuy}{\ensuremath{c_\g{grid,buy}}\xspace}

\newcommand{\cgridsell}{\ensuremath{c_\g{grid,sell}}\xspace}

\newcommand{\cgridpeak}{\ensuremath{c_\g{grid,peak}}\xspace}

\newcommand{\Cthi}[1][]{
	\ifthenelse{ \equal{#1}{} }
	{\ensuremath{C_{\g{th},i}}\xspace}
	{\ensuremath{C_{\g{th},{#1}}}\xspace}
}

\newcommand{\Hai}[1][]{
	\ifthenelse{ \equal{#1}{} }
	{\ensuremath{H_{\g{air},i}}\xspace}
	{\ensuremath{H_{\g{air},{#1}}}\xspace}
}

\renewcommand{\sc}{\ensuremath{s_\g{c}}\xspace}

\newcommand{\Pgrid}{\ensuremath{P_\g{grid}}\xspace}

\newcommand{\Pgridpeak}{\ensuremath{P_\g{grid,peak}}\xspace}
\newcommand{\Pgridpeakk}{\ensuremath{P_\g{grid,peak}(k)}\xspace}

\newcommand{\Qcooli}[1][]{
	\ifthenelse{ \equal{#1}{} }
	{\ensuremath{\dot{Q}_{\g{cool,}i}}\xspace}
	{\ensuremath{\dot{Q}_\g{cool,{#1}}}\xspace}
}

\newcommand{\Qheati}[1][]{
	\ifthenelse{ \equal{#1}{} }
	{\ensuremath{\dot{Q}_{\g{heat,}i}}\xspace}
	{\ensuremath{\dot{Q}_\g{heat,{#1}}}\xspace}
}

\newcommand{\Qotheri}[1][]{
	\ifthenelse{ \equal{#1}{} }
	{\ensuremath{\dot{Q}_{\g{other,}i}}\xspace}
	{\ensuremath{\dot{Q}_{\g{other,{#1}}}}\xspace}
}

\newcommand{\thetabi}[1][]{
	\ifthenelse{ \equal{#1}{} }
	{\ensuremath{\vartheta_{\g{b},i}}\xspace}
	{\ensuremath{\vartheta_{\g{b,{#1}}}}\xspace}
}

\newcommand{\thetait}[1][]{
	\ifthenelse{ \equal{#1}{} }
	{\ensuremath{\vartheta_{i}(t)}\xspace}
	{\ensuremath{\vartheta_{#1}(t)}\xspace}
}

\newcommand{\Qtotit}[1][]{
	\ifthenelse{ \equal{#1}{} }
	{\ensuremath{\dot{Q}_{\g{tot,}i}}\xspace}
	{\ensuremath{\dot{Q}_{\g{tot,}{#1}}}\xspace}
} 

\newcommand{\Tsamp}{\ensuremath{T_\g{s}}\xspace}
	\newcommand{\Tsampseq}{\ensuremath{\boldsymbol{T}_\g{s}}\xspace}
\newcommand{\Tsampk}{\ensuremath{\Tsamp(k)}\xspace}

\newcommand{\Pdemk}{\ensuremath{P_\g{dem}(k)}\xspace}

\newcommand{\Imat}[2][]{
	\ifthenelse{ \equal{#1}{} }
	{\ensuremath{\boldsymbol{I}}\xspace}
	{\ensuremath{\boldsymbol{I}_{{#1}\times {#2}} }\xspace}
}
\newcommand{\Zeromat}[2][]{
	\ifthenelse{ \equal{#1}{} }
	{\ensuremath{\boldsymbol{0}}\xspace}
	{\ensuremath{\boldsymbol{0}_{{#1}\times {#2}} }\xspace}
}
\newcommand{\Onemat}[2][]{
	\ifthenelse{ \equal{#1}{} }
	{\ensuremath{\boldsymbol{1}}\xspace}
	{\ensuremath{\boldsymbol{1}_{{#1}\times {#2}} }\xspace}
}

\newcommand{\Eevi}[1][]{
	\ifthenelse{ \equal{#1}{} }
	{\ensuremath{E_{\g{EV},i}}\xspace}
	{\ensuremath{E_{\g{EV,{#1}}}}\xspace}
}
\newcommand{\Eevarri}[1][]{
	\ifthenelse{ \equal{#1}{} }
	{\ensuremath{E_{\g{EV,arr},i}}\xspace}
	{\ensuremath{E_{\g{EV,arr,{#1}}}}\xspace}
}
\newcommand{\Eevdepi}[1][]{
	\ifthenelse{ \equal{#1}{} }
	{\ensuremath{E_{\g{EV,dep},i}}\xspace}
	{\ensuremath{E_{\g{EV,dep,{#1}}}}\xspace}
}
\newcommand{\Eevmini}[1][]{
	\ifthenelse{ \equal{#1}{} }
	{\ensuremath{E_{\g{EV,min},i}}\xspace}
	{\ensuremath{E_{\g{EV,min,{#1}}}}\xspace}
}

\newcommand{\Cevi}[1][]{
	\ifthenelse{ \equal{#1}{} }
	{\ensuremath{C_{\g{EV},i}}\xspace}
	{\ensuremath{C_{\g{EV,{#1}}}}\xspace}
}
\newcommand{\Cevarri}[1][]{
	\ifthenelse{ \equal{#1}{} }
	{\ensuremath{C_{\g{EV,arr},i}}\xspace}
	{\ensuremath{C_{\g{EV,arr,{#1}}}}\xspace}
}
\newcommand{\Cevdepi}[1][]{
	\ifthenelse{ \equal{#1}{} }
	{\ensuremath{C_{\g{EV,dep},i}}\xspace}
	{\ensuremath{C_{\g{EV,dep,{#1}}}}\xspace}
}

\newcommand{\Pevi}[1][]{
	\ifthenelse{ \equal{#1}{} }
	{\ensuremath{P_{\g{EV},i}}\xspace}
	{\ensuremath{P_{\g{EV,{#1}}}}\xspace}
}
\newcommand{\Pevmaxi}[1][]{
	\ifthenelse{ \equal{#1}{} }
	{\ensuremath{P_{\g{EV,max},i}}\xspace}
	{\ensuremath{P_{\g{EV,max,{#1}}}}\xspace}
}

\newcommand{\Devi}[1][]{
	\ifthenelse{ \equal{#1}{} }
	{\ensuremath{D_\g{i}}\xspace}
	{\ensuremath{D_{\g{#1}}}\xspace}
}

\newcommand{\indlpmp}{\g{LP}\rightarrow\g{MP}}
\newcommand{\indcomplpmp}{\g{comp,}\indlpmp}

\newcommand{\mlp}{\ensuremath{m_\g{LP}}\xspace}
\newcommand{\mmp}{\ensuremath{m_\g{MP}}\xspace}

\newcommand{\mmpi}[1][]{
	\ifthenelse{ \equal{#1}{} }
	{\ensuremath{m_{\g{MP},i}}\xspace}
	{\ensuremath{m_{\g{MP},{#1}}}\xspace}
}
\newcommand{\mmpimin}{\ensuremath{m_{\g{MP},i,\g{min}}}\xspace}

\newcommand{\mdotely}{\ensuremath{\dot{m}_\g{ely}}\xspace}
\newcommand{\Pely}{\ensuremath{P_\g{ely}}\xspace}

\newcommand{\mdotlpmp}{\ensuremath{\dot{m}_{\indlpmp}}\xspace}
\newcommand{\mdotfuelmp}{\ensuremath{\dot{m}_{\g{MP}\rightarrow\g{Fuel}}}\xspace}
\newcommand{\mdotfuelmpseq}{\ensuremath{\dot{\boldsymbol{m}}_{\g{MP}\rightarrow\g{Fuel}}}\xspace}

\newcommand{\Pcomplpmp}{\ensuremath{P_{\indcomplpmp}}\xspace}

\newcommand{\zfuelmismatch}{\ensuremath{z_\g{fuel}}\xspace}

\newcommand{\belyon}{\ensuremath{b_\g{ely,on}}\xspace}
\newcommand{\btildeelyon}{\ensuremath{\tilde{b}_\g{ely,on}}\xspace}
\newcommand{\bcomplpmp}{\ensuremath{b_{\indcomplpmp}}\xspace}
\newcommand{\bcomppr}{\ensuremath{b_\g{comp,PR}}\xspace}

\newcommand{\cgridco}{\ensuremath{c_\g{grid,\CO}}\xspace}

\newcommand{\cgriddiff}{\ensuremath{c_\g{grid,diff}}\xspace}

\newcommand{\cfueldemandmis}{\ensuremath{c_\g{fuel demand}^\g{mismatch}}\xspace}
\newcommand{\cstartup}{\ensuremath{c_\g{startup}}\xspace}

\newcommand{\belyready}{\ensuremath{b^{\g{measured}}_\g{ely,ready}}\xspace}
\newcommand{\bconsiderpr}{\ensuremath{b_\g{consider PR}}\xspace}
\newcommand{\bconsiderprseq}{\ensuremath{\boldsymbol{b}_\g{consider PR}}\xspace}

\newcommand{\mdotfueldemand}{\ensuremath{\dot{m}_\g{fuel demand}}\xspace}

\newcommand{\tpr}{\ensuremath{t_\g{PR}}\xspace}

\newcommand{\Ppv}{\ensuremath{P_\g{PV}}\xspace}

\newcommand{\mlpmin}{\ensuremath{m_\g{LP,min}}\xspace}
\newcommand{\mlpmax}{\ensuremath{m_\g{LP,max}}\xspace}
\newcommand{\mlpminsoft}{\ensuremath{m_\g{LP,min}^\g{soft}}\xspace}

\newcommand{\mmpmin}{\ensuremath{m_\g{MP,min}}\xspace}
\newcommand{\mmpmax}{\ensuremath{m_\g{MP,max}}\xspace}
\newcommand{\mmpminsoft}{\ensuremath{m_\g{MP,min}^\g{soft}}\xspace}

\newcommand{\Pelymin}{\ensuremath{P_\g{ely,min}}\xspace}
\newcommand{\Pelymax}{\ensuremath{P_\g{ely,max}}\xspace}

\newcommand{\Pcomppr}{\ensuremath{P_\g{comp,PR}}\xspace}
\newcommand{\Pcompmax}{\ensuremath{P_\g{comp,max}}\xspace}

\newcommand{\mdotprnom}{\ensuremath{\dot{m}_\g{PR,nominal}}\xspace}

\newcommand{\thetampi}[1][]{
	\ifthenelse{ \equal{#1}{} }
	{\ensuremath{\vartheta_{\g{MP},i}}\xspace}
	{\ensuremath{\vartheta_{\g{MP},{#1}}}\xspace}
}

\newcommand{\plp}{\ensuremath{p_\g{LP}}\xspace}
\newcommand{\pmp}{\ensuremath{p_\g{MP}}\xspace}

\newcommand{\pmpi}[1][]{
	\ifthenelse{ \equal{#1}{} }
	{\ensuremath{p_{\g{MP},i}}\xspace}
	{\ensuremath{p_{\g{MP},{#1}}}\xspace}
}

\newcommand{\mseci}[1][]{
	\ifthenelse{ \equal{#1}{} }
	{\ensuremath{m_{\g{MP,sec},i}}\xspace}
	{\ensuremath{m_{\g{MP,sec},{#1}}}\xspace}
}

\newcommand{\mdotlpmpsamples}{\ensuremath{\dot{m}^\g{\indlpmp}_{\g{s}}}\xspace}

\newcommand{\mlpsamples}{\ensuremath{m_{\g{LP,s}}}\xspace}

\newcommand{\Jsclp}{\ensuremath{J_{\g{SC,LP}}}\xspace}
\newcommand{\Jscmp}{\ensuremath{J_{\g{SC,MP}}}\xspace}

\newcommand{\Jmongrid}{\ensuremath{J_\g{mon}^\g{grid}}\xspace}

\newcommand{\Jmonpeak}{\ensuremath{J_\g{mon}^\g{peak}}\xspace}
	\newcommand{\Jmonpeakprev}{\ensuremath{J_\g{mon,peak}^{\g{previous}}}\xspace}
\newcommand{\Jmontotal}{\ensuremath{J_\g{mon,total}}\xspace}

\newcommand{\ctildepeak}{\ensuremath{\tilde{c}_\g{peak}^\g{max}}\xspace}

\newcommand{\Jstartup}{\ensuremath{J_\g{startup}}\xspace}
\newcommand{\toggleonoff}{\ensuremath{\delta_\g{on-off}}\xspace}

\newcommand{\Juser}{\ensuremath{J_\g{user}}\xspace}
\newcommand{\Jco}{\ensuremath{J_\g{\CO}}\xspace}

\newcommand{\Jtotal}{\ensuremath{J_\g{total}}\xspace}

\newcommand{\Nsessions}{\ensuremath{N_{\g{sessions}}}\xspace}
\newcommand{\tarr}{\ensuremath{t_{\g{arrival}}}\xspace}
\newcommand{\mhtwo}{\ensuremath{m_{\g{H_2}}}\xspace}
\newcommand{\mhtwotilde}{\ensuremath{\tilde{m}_{\g{H_2}}}\xspace}
\newcommand{\nul}{\ensuremath{(0)}\xspace}

\newcommand{\Pavailable}{\ensuremath{P_\g{available}}\xspace}
	\newcommand{\Pavailablepeak}{\ensuremath{P_\g{available}^\g{Peak}}\xspace}
	\newcommand{\Pavailableexcess}{\ensuremath{P_\g{available}^\g{Excess}}\xspace}
\newcommand{\melymin}{\ensuremath{m_\g{ely,min}}\xspace}
\newcommand{\melymax}{\ensuremath{m_\g{ely,max}}\xspace}
\newcommand{\Pres}{\ensuremath{P_\g{res}}\xspace}
\newcommand{\Pmax}{\ensuremath{P_\g{max}}\xspace}

\newcommand{\mmplimalloc}{\ensuremath{m_\g{MP,lim}^\g{alloc}}}
\newcommand{\mmplimallocseq}{\ensuremath{\boldsymbol{m}_\g{MP,lim}^\g{alloc}}}
\newcommand{\zlimalloc}{\ensuremath{z_\g{lim,alloc}}}
\newcommand{\kcutoff}{\ensuremath{n_\g{cutoff}}}

\newcommand{\useqadj}{\ensuremath{\boldsymbol{\uvek_\g{adj}}}\xspace}

\newcommand{\mprtemp}{\ensuremath{m_\g{PR,temp}}\xspace}
\newcommand{\mprplan}{\ensuremath{m_\g{PR,plan}}\xspace}
\newcommand{\mpravailable}{\ensuremath{m_\g{PR,available}}\xspace}
\newcommand{\mprtomove}{\ensuremath{m_\g{PR,to\,move}}\xspace}

\newcommand{\letterhighersec}{\ensuremath{h}\xspace}
\newcommand{\letterlowersec}{\ensuremath{\ell}\xspace}

\newcommand{\mfilled}{\ensuremath{m_\g{filled}}\xspace}
\newcommand{\mfilledhighersec}{\ensuremath{m_{\g{filled},\letterhighersec}}\xspace}
\newcommand{\mfilledlowersec}{\ensuremath{m_{\g{filled},\letterlowersec}}\xspace}
\newcommand{\mtofill}{\ensuremath{m_\g{to\,fill}}\xspace}

\newcommand{\mtofuel}{\ensuremath{m_\g{to\,fuel}}\xspace}
\newcommand{\mdotfuelreal}{\ensuremath{\dot{m}_\g{fuel,real}}\xspace}
\newcommand{\mdotfuelusei}{\ensuremath{\dot{m}_{\g{fuel,use},i}}\xspace}

\newcommand{\mdotfuelmismatch}{\ensuremath{\dot{m}_{\g{fuel,mismatch}}}\xspace}
\newcommand{\mdotfuelmismatchseq}{\ensuremath{\dot{\boldsymbol{m}}_{\g{fuel,mismatch}}}\xspace}

\newcommand{\nfirstmismatch}{\ensuremath{n_{\g{fm}}}\xspace}
\newcommand{\mavaforpr}{\ensuremath{m_\g{available\,for\,PR}}\xspace}
\newcommand{\mmovewithpr}{\ensuremath{m_\g{move\,with\,PR}}\xspace}
\newcommand{\mmppossible}{\ensuremath{m_\g{MP,possible}}\xspace}
\newcommand{\mmppossibleseq}{\ensuremath{\boldsymbol{m}_\g{MP,possible}}\xspace}
\newcommand{\mmpthreefiftylimit}{\ensuremath{m_\g{MP,350\,\barpressure\,limit}}\xspace}

\let\st\relax
\DeclareMathOperator*{\st}{s.t.}
\DeclareMathOperator*{\AND}{AND}

\newcommand{\mylinevec}[1]{
	\begingroup
	\setlength\arraycolsep{2pt}
	\begin{pmatrix}
		#1
	\end{pmatrix}
	\endgroup
	\xspace
}

\newcommand{\pluseq}{\mathrel{+}=}
\newcommand{\minuseq}{\mathrel{-}=}

%% file: common/macros_common.tex
\newcommand{\eg}{e.\thinspace{}g.\@\xspace}
\newcommand{\ie}{i.\thinspace{}e.\@\xspace}

\newcommand{\pagerefh}[1]{\hyperref[#1]{page~\pageref*{#1}}}
\newcommand{\secref}[1]{\hyperref[#1]{Section~\ref*{#1}}}
\newcommand{\chapref}[1]{\hyperref[#1]{Chapter~\ref*{#1}}}
\newcommand{\figref}[1]{Figure~\ref{#1}}
\newcommand{\tabref}[1]{\hyperref[#1]{Table\ \ref*{#1}}}

\newcommand{\algoref}[1]{Algorithm~\ref{#1}}
\newcommand{\g}{\mathrm}

\newcommand{\kW}{\ensuremath{\g{kW}}\xspace}
\newcommand{\kWh}{\ensuremath{\g{kWh}}\xspace}

\newcommand{\kg}{\ensuremath{\g{kg}}\xspace}
\newcommand{\barpressure}{\ensuremath{\g{bar}}\xspace}

\newcommand{\seconds}{\ensuremath{\,\g{s}}\xspace}
\newcommand{\minutes}{\ensuremath{\,\g{min}}\xspace}
\newcommand{\hours}  {\ensuremath{\,\g{h}}\xspace}

\newcommand\MEUR[1]{\if@EURleft\text{\euro}\,\fi#1\if@EURleft\else\,\text{\euro}\fi}
\newcommand{\meuro}{\text{\euro}}

\newcommand{\meuroperkwnice}{\ensuremath{\nicefrac{\meuro}{\kW}}\xspace}

\newcommand{\meuroperkwhnice}{\ensuremath{\nicefrac{\meuro}{\kWh}}\xspace}

\newcommand{\CO}{ \ensuremath{ \g{CO}_2 }\xspace}

\newcommand{\percent}{\,\%\xspace}

%% file: common/glossary_entries.tex
\newacronym{bem}{BEM}{building energy management}
\newacronym{res}{RES}{renewable energy source}
\newacronym{mpc}{MPC}{Model Predictive Control}
\newacronym{moo}{MOO}{multi-objective optimization}
\newacronym{ev}{EV}{electric vehicle}
\newacronym{bps}{BPS}{building performance software}
\newacronym{ode}{ODE}{ordinary differential (or difference) equation}
\newacronym{hri}{HRI}{Honda Research Institute}
\newacronym{hvac}{HVAC}{heating, ventilation, and air conditioning}
\newacronym{chp}{CHP}{combined heat and power plant}
\newacronym{pv}{PV}{photovoltaic}
\newacronym{empc}{EMPC}{Economic Model Predictive Control}
\newacronym{der}{DER}{distributed energy resource}
\newacronym{ess}{ESS}{energy storage system}
\newacronym{tes}{TES}{thermal energy storage}
\newacronym{ocp}{OCP}{optimal control problem}
\newacronym{qp}{QP}{quadratic programming}
\newacronym{nlp}{NLP}{nonlinear programming}
\newacronym{milp}{MILP}{mixed-integer linear programming}
\newacronym{minlp}{MINLP}{mixed-integer nonlinear programming}
\newacronym{dm}{DM}{decision maker}
\newacronym{awds}{AWDS}{adaptive weight determination scheme}
\newacronym{nbi}{NBI}{normal boundary intersection}
\newacronym{chim}{CHIM}{convex hull of individual minima}
\newacronym{fpbi}{FPBI}{focus point boundary intersection}
\newacronym{cup}{CUP}{closest to Utopia point}
\newacronym{aep}{AEP}{angle to the extreme points}
\newacronym{atn}{ATN}{angle to the neighbor points}
\newacronym{ci}{CI}{carbon intensity}
\newacronym{hl}{HL}{higher level}
\newacronym{ll}{LL}{lower level}

\newacronym[\glslongpluralkey={Gaussian Processes}]{gp}{GP}{Gaussian Process}
\newacronym{bepst}{BEPST}{building energy performance simulation tool}
\newacronym{ann}{ANN}{artificial neural network}
\newacronym{rnn}{RNN}{recurrent neural network}
\newacronym{ems}{EMS}{energy management system}
\newacronym{svr}{SVR}{support vector regression}
\newacronym{dr}{DR}{demand response}
\newacronym{ga}{GA}{Genetic Algorithm}
\newacronym{rc}{RC}{Resistor-Capacitor}
\newacronym{rbf}{RBF}{radial basis function}
\newacronym{pmv}{PMV}{predicted mean vote}
\newacronym{ekf}{EKF}{extended Kalman filter}
\newacronym{ml}{ML}{machine learning}
\newacronym{kpi}{KPI}{key performance indicator}

\newacronym{sil}{SiL}{software-in-the-loop}
\newacronym{fmu}{FMU}{functional mock-up unit}

\newacronym{fcv}{FCV}{fuel cell vehicle}
\newacronym{hrs}{HRS}{hydrogen refueling station}
\newacronym{pem}{PEM}{proton exchange membrane}
\newacronym{pwla}{PWLA}{piece-wise linear approximation}
\newacronym{mp}{MP}{medium pressure}
\newacronym{lp}{LP}{low pressure}
\newacronym{pr}{PR}{pressure recovery}
\newacronym{dpp}{DPP}{disciplined parametrized programming}

\newacronym{rbcp}{RBC Peak}{RBC Peak}
\newacronym{rbce}{RBC Excess}{RBC Excess}

%% file: inc/01_intro.tex
\section{Introduction}
\label{sec:intro}

\subsection{Motivation} 

The push for decarbonization of the global economy has spurred an increasing adoption of renewable energy sources like wind and solar power. 
However, the variability of these sources poses challenges for integration into the centralized public power grid. 
Therefore, local microgrids will become more important to support the decentralization of the power grid. 
Hydrogen emerges as a compelling solution for long-term energy storage due to its high energy density and the possibility of emission-free production.
In addition, fossil fuels will have to be replaced in the transportation sector. 
While battery electric vehicles are likely to represent the bulk of the private transportation sector, the number \glspl{fcv} is expected to increase, too, with different projections from 582,000 units in 2030 worldwide \cite{inci2021review} up to over 10,000,000 \cite{samsun2022deployment}. 
In combination with the above mentioned decentralization, this necessitates the intelligent production and storage of hydrogen, possibly in form of local \glspl{ems}. 

As part of a research project, Honda Europe R\&D (Deutschland) GmbH is currently installing a hydrogen production system including a \gls{hrs} in Offenbach, Germany. 
It will be connected to the currently existing energy infrastructure \cite{stadie2021v2b} and consists of an electrolyzer, a compressor, a \gls{lp} storage tank and six \gls{mp} storage tanks. 
The \gls{mp} tanks are divided into two sections, and their physical coupling during the filling and removal of hydrogen follows complex rules. 
In this paper, we present an approach on how an \gls{ems} based on \gls{mpc} can be developed for the planned system despite its complexity. 
	
\subsection{Literature Review}

\gls{mpc} is regularly being considered for hydrogen systems. 
However, while many works extensively consider the modeling of the hydrogen production part, \eg the electrolyzer's efficiency curve or security constraints \cite{huang2023economic}, the hydrogen's pressure when stored is usually neglected. 
In \cite{cardona2023model}, \gls{mpc} is used for an on-site \gls{hrs} which is set to be built in Zaragoza, Spain. 
The plant consists of multiple electrolyzers, compressors and storage tanks and should serve both light duty vehicles (700\,\barpressure) and heavy duty vehicles (350\,\barpressure). 
While it has 3 cascaded tanks to serve the dispenser of the heavy duty vehicles with a maximum of 500\,\barpressure, the \gls{mpc} does not respect the actual pressure levels of the tanks. 
Namely, neither the mass flow rates vary, nor is there a distinction if the tank pressure is high enough to refill the vehicle completely.
In \cite{abdelghany2022two}, a two-layered \gls{mpc} strategy is developed to control a wind-powered hydrogen-based microgrid. 
The higher level \gls{mpc} schedules the hydrogen production of an electrolyzer for a time horizon of 24\hours while respecting the demand from \glspl{fcv} and possible grid loads. 
The lower level \gls{mpc} is then supposed to track the production setpoints in real-time with a time horizon of 1\hours, while additionally respecting the electrolyzer's status (warm vs. cold) in form of an automaton. 
However, the hydrogen storage's pressure is neglected completely. 
This neglection in combination with \gls{mpc} is most likely due to the complexity of modeling the fueling process with all relevant parameters, \eg temperature and pressure.
Usually, these processes are modeled with modeling languages like Modelica \cite{kuroki2018dynamic,rothuizen2013optimization}, which support necessary mathematical descriptions like partial differential equations. 
Unfortunately, the resulting models are not well suited for \gls{mpc} due to the complex solution of the resulting \gls{ocp}. 
While it could be solved using evolutionary algorithms, mixed-integer programming has been shown superior for a similar problem \cite{singh2022investigating, ishihara2020optimizing}. 
Thus, the hydrogen's pressure shall be respected in the \gls{mpc} more implicitly. 

One option would be to do so by tracking externally provided setpoints. 
This is common practice in \gls{mpc} to optimize objectives which are not (or cannot be) explicitly represented.  
In this case, the setpoint might be derived from an (arbitrarily complex) heuristic. 
For example, in \cite{cai2015generalized}, a heuristic strategy is used to achieve near-optimal control of direct expansion units of air conditioning systems by adjusting the supply air temperature set-point in dependence of zone conditions and coil status. 
Note that by providing external (steady-state) setpoints, usually the optimality in regard to the original objectives is lost, which has been the main motivation for the development of economic \gls{mpc} \cite{amrit2011economic}. 

In general, the \gls{mpc} incorporates the model of a system as constraints in the \gls{ocp}, \eg ordinary difference equations describing the model dynamics in the form of $\xkp = f(\xk, \uk)$ where \xvek is the state vector and \uvek the input vector. 
However, if a part of the system dynamics is too complex to respect it \textit{explicitly} in this form within the \gls{ocp}, it is possible to do so \textit{implicitly} by additional constraints. 
For example, in \cite{nabais2013constrained}, \gls{mpc} is used to control the cargo transportation in intermodal hubs.
Since the optimization of the transport modal split would be too complex, a heuristic is used. 
Then, the desired split is incorporated in form of a terminal constraint, which is relaxed to ensure feasibility.

The plant considered in this study has two processes which are too complex to be explicility considered in an \gls{ocp}, namely the order in which the 6 \gls{mp} tanks are fueled and emptied depending on their individual pressure levels, and the \textit{pressure recovery} functionality, which shifts hydrogen between 2 \gls{mp} sections.
Thus, as our main contribution, we propose a novel control strategy to implicitly incorporate these processes. 
The \gls{mpc} solves the \gls{ocp} first neglecting the pressure recovery and individual pressure levels. 
Then, an alternative (more complex) model is used by the so-called \textit{allocator} to validate whether the \gls{mpc}'s plan is feasible or not, \ie whether all \glspl{fcv} can be fueled as planned. 
If not, the allocator sets additional constraints for the \gls{ocp} to ensure the fueling success and the \gls{mpc} runs a second time. 
Through this approach, we can respect the necessity of sufficiently high storage pressure levels in the fueling process, which is otherwise neglected in the literature. 
Additionally, we also respect the storage pressure in the calculation of the compressor's mass flow and power consumption while maintaining convexity by \glspl{pwla}. 

The rest of the paper is structured as follows. 
A more detailed overview of the planned hydrogen system is given in \secref{sec:system_overview}, 
followed by the description of our proposed control approach in \secref{sec:control_approach}. 
A simulation study showing its advantages over rule-based controllers is presented in \secref{sec:simulation_study}, followed by a brief discussion and outlook in \secref{sec:conclusion}. 

%% file: inc/02_system_overview.tex
\section{System Overview}
\label{sec:system_overview}

\begin{figure*}[htb]
	\centering
	\textbf{\small Hydrogen Plant Overview}\par 
	\vspace{2mm}
	\includegraphics{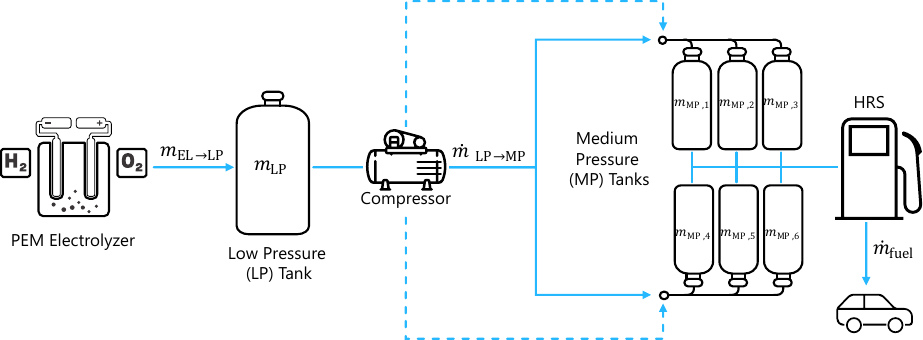}	
	\caption{
		The electrolyzer directly fills the \gls{lp} tank of max. $11\,\kg / 30\,\barpressure$ capacity. 
		A compressor is then used to transfer the hydrogen to 6 individual \gls{mp} tanks with max. $43.33\,\kg  / 450\,\barpressure$ capacity each, which are organized in 2 sections with 3 tanks each. 
		The compressor can also be used to shift hydrogen between these sections (\textit{pressure recovery}). 
		The \gls{hrs} (dispenser) is connected to the \gls{mp} tanks and used to refuel \glspl{fcv} with $350\,\barpressure$. 
	}
	\label{fig:system_overview}
\end{figure*}

An overview of the hydrogen plant is depicted in \figref{fig:system_overview}. 
In the following subsections, we give a brief description of the individual entities and how they are modeled in the \gls{mpc}. 
We use the standard \gls{mpc} notation $r\nk$, which refers to the value of $r(k+n)$ predicted at time step $k$. 
Note that we use varying step sizes $\Tsamp(n|k)$ for different $n$ within the 7-day prediction horizon of the \gls{mpc}, which has to be respected in the following.

\subsection{Electrolyzer}
\label{sec:system_overview-electrolyzer}

The core of the plant is a \gls{pem} electrolyzer with a maximum power consumption of $\Pelymax = 225\,\kW$.  
It also has a minimum power consumption of $\Pelymin = 70\,\kW$. 
Thus, its on/off status has to be considered. 
Additionally, it has a warm up time of approx. 15\minutes, which we model by two different binary variables. 
$\belyon \in \{0, 1\}$ denotes the "on" signal sent by the controller, 
and the auxiliary variable $\btildeelyon \in \{0, 1\}$ denotes whether it is ready to produce hydrogen. 
The power limits are then expressed by 
\begin{IEEEeqnarray}{rCl}
	\btildeelyon\k \cdot \Pelymin & \leq & \Pely\k, \label{eq:b_tilde_1} \\ 
	\Pely\k & \leq & \btildeelyon\k \cdot \Pelymax. \label{eq:b_tilde_2} 
\end{IEEEeqnarray}
Additionally, \btildeelyon is constrained by logical AND conditions \cite[\S7.7]{bisschop2006aimms} on the previous values of \belyon.
Due to the varying step sizes, every $\btildeelyon\nk$ in the horizon may have a different number of constraints. 
However, to reduce the number of binary decision variables and because later in the horizon, the time steps are bigger than the actual start up constraint time, we do not consider the startup constraints in the entire horizon. 
The nonlinear mapping of the hydrogen output in dependence of the consumed electrical power, $\mdotely = f_\g{ely}(\Pely)$,  is formulated as a one-dimensional \gls{pwla} \cite{dambrosio2010piecewise}, which results in additional binary variables and constraints. 
For this study, we use measurement points taken from \cite{kopp2018strommarktseitige}.

\subsection{Tanks}
\label{sec:system_overview-tanks}

The \gls{lp} tank has a maximum capacity of $\mlpmax = 11\,\kg$ at $30\,\barpressure$ and is directly filled by the electrolyzer. 
The 6 \gls{mp} tanks have a maximum capacity of $\approx 43.33\,\kg$ at $450\,\barpressure$ each. 
However, due to their complex couplings, they are modelled as a single aggregated tank with $\mmpmax = 260\,\kg$. 
Additionally, lower bounds $\mlpmin = 0.5\,\kg$ and $\mmpmin = 60\,\kg$ apply, defined by the constraints 
\begin{IEEEeqnarray}{rCcCl}
	\mlpmin & \leq & \mlp\k & \leq & \mlpmax \label{eq:m_lp_box}, \\ 
	\mmpmin & \leq & \mmp\k & \leq & \mmpmax \label{eq:m_mp_box}.  
\end{IEEEeqnarray}
The relationship between a tank's pressure and its mass is slightly nonlinear (concave) and temperature dependent. 
However, for simplicity, in the following we assume $\plp \propto \mlp$ and $\pmp \propto \mmp$. 

The 6 \gls{mp} tanks are organized in 2 sections with 3 tanks each. 
The filling of the \gls{mp} tanks (from the \gls{lp} tank) is done section-wise. 
Namely, the section with the higher average pressure is filled first. 
Within this section, the tank with the lowest pressure is filled until it has the same pressure as the second lowest tank. 
Then, these two are filled simultaneously until they have the same pressure as the third, from where on all three are filled simultaneously. 
When the section hits its upper limit, the other section is being filled according to the same rules.

With the pressure recovery function, hydrogen can also be shifted between the two sections. 
Thereby, hydrogen is taken from the section with the lower average pressure, and in there from the tank with the lowest pressure, until it hits the lower limit. 
Then, the next lowest tank (of the same section) is used. 
The filling of the individual tanks of the other section follows the same rules as described above. 
 
Note that modeling these heuristics explicitly would lead to a non-convex optimization problem. 
Thus, they are considered implicitly in the proposed allocator approach, see \secref{sec:control_approach-allocator}. 

\subsection{Compressor}
\label{sec:system_overview-compressor}

The compressor can be used for either transferring hydrogen from the \gls{lp} tank to the \gls{mp} tanks, or for the pressure recovery function. 
Thus, we introduce binary variables for both modes, \bcomplpmp and \bcomppr, with the constraint 
\begin{IEEEeqnarray}{rCl}
	\bcomplpmp\k + \bcomppr\k & \leq & 1. \label{eq:bcomp_logic}
\end{IEEEeqnarray}
In either case, the compressor's power cannot be modulated.
For $\bcomplpmp = 1$, the resulting mass flow \mdotlpmp depends on the input pressure, \ie on \mlp. 
This is respected by a one-dimensional \gls{pwla} \cite{dambrosio2010piecewise}, using the sample points $\mlpsamples~\hat{=}~ (0, 20, 90)\,\barpressure$ and $\mdotlpmpsamples = (0.2, 4.2, 18.0)\,\frac{\kg}{\hours}$. 
The necessary power \Pcomplpmp additionally also significantly depends on the output pressure, \ie \mmp. 
Thus, a two-dimensional \gls{pwla} based on the triangle method is used \cite{dambrosio2010piecewise}. 
For both cases, the resulting constraints are omitted for brevity.

\subsection{Hydrogen Refueling Station (HRS)}
\label{sec:system_overview-fueling_station}

For the refueling of \glspl{fcv}, the uncontrollable demand is modeled as a disturbance \mdotfueldemand in $\frac{\kg}{\hours}$. 
However, the provided hydrogen \mdotfuelmp may deviate, which is expressed as the constraint 
\begin{IEEEeqnarray}{rCl}
	\mdotfueldemand\k & = & \mdotfuelmp\k + \zfuelmismatch\k \label{eq:slack_fuel} 
\end{IEEEeqnarray}
with a slack variable \zfuelmismatch, which is punished in the objective function, see \eqref{eq:Juser}. 
The refueling process is done tank-wise, \ie the \gls{mp} tank with the lowest possible pressure (\ie above the car's tank pressure) is always used. 
Once the \gls{mp} tank and the car's tank are at the same pressure level, the next higher \gls{mp} tank is used. 
To avoid modeling the car tank, we use the \gls{mp} tank with the lowest pressure above $350\,\barpressure$. 

%% file: inc/03_control_approach.tex
\section{Control Approach}
\label{sec:control_approach}

To develop the \gls{mpc} for the above described plant, in this section we first complement its model with the main dynamics, then define objectives to optimize and formulate the resulting \gls{ocp}. 
Lastly, we present the novel allocator approach in \secref{sec:control_approach-allocator} as our main contribution. 

\subsection{Plant Model}
\label{sec:control_approach-plant_model}

The discretized tank masses in $\kg$ can be described by the ordinary difference equations 
\begin{IEEEeqnarray}{rcl}
	\mlp\kpo & = & \mlp\k\!+\!(\mdotely\k\! - \!\mdotlpmp\k) \cdot \Tsampk, \IEEEeqnarraynumspace \label{eq:dyn_lp} \\
	\mmp\kpo & = & \IEEEnonumber \\
	\IEEEeqnarraymulticol{3}{c}{
		\mmp\k + (\mdotlpmp\k - \mdotfuelmp\k\big) \cdot\Tsampk, 
	}
	\label{eq:dyn_mp}
\end{IEEEeqnarray}
where \mdotlpmp is the compressor's and \mdotfuelmp the \gls{hrs}'s mass flow in $\frac{\kg}{\hours}$. 

The electrical balance equation is given by 
\begin{IEEEeqnarray}{rCl}
	0 & = & \Pgrid\k + \Pdemk + \Ppv\k - \Pely\k \IEEEnonumber \\
	&& -\>  \bcomplpmp\k \cdot \Pcomplpmp\k  \IEEEnonumber \\ 
	&& -\> \bcomppr\k \cdot \Pcomppr,   \label{eq:bal_eq} 
\end{IEEEeqnarray}
where we assume a constant value for \Pcomppr.

\subsection{Objectives}
\label{sec:control_approach-objectives}

\subsubsection{Soft Constraints on Tank Limits}

For both the LP and MP tanks, 
increased lower limits ($\mlpminsoft = 7\,\kg > \mlpmin$, $\mmpminsoft = 151.9\,\kg > \mmpmin$)  
are enforced using soft constraints. 
Instead of a relaxed version of constraints \eqref{eq:m_lp_box} and \eqref{eq:m_mp_box}, we introduce the objective costs
\begin{IEEEeqnarray}{rcl}
	\Jsclp\k & = & 
	0.1 \! \cdot \! 
	\sumnNppo \! \max\!\left(0,\mlpminsoft \! - \! \mlp\nk \right)\!\cdot\!\Tsamp\nk, \IEEEnonumber \\
	\label{eq:sc_lp} \\
	\Jscmp\k & = & 
	0.1 \! \cdot \! 
	\sumnNppo\! \max\!\left(0,\mmpminsoft \! - \! \mmp\nk\right)\!\cdot\!\Tsamp\nk. \IEEEnonumber \\ 
	\label{eq:sc_mp} 
\end{IEEEeqnarray}
\mlpminsoft is chosen such that an input pressure of at least $20\,\barpressure$ for the compressor is ensured.  
\mmpminsoft is chosen such that there is enough hydrogen to have one section (\ie 3 tanks) above $350\,\barpressure$, plus additional $4\,\kg$ hydrogen to refuel a car at any time. 

\subsubsection{Monetary Costs}
For the grid costs, different prices $\cgridbuy = 0.144\,\meuroperkwhnice$ for buying and $\cgridsell = 0.07\,\meuroperkwhnice$ for selling are assumed. 
To formulate them not only convexly, but also following \gls{dpp}\footnote{See \url{https://www.cvxpy.org/tutorial/advanced/index.html\#disciplined-parametrized-programming}} rules, we use an auxiliary parameter $\cgriddiff = \cgridbuy - \cgridsell > 0$, with which the costs are described as  
\begin{IEEEeqnarray}{rcrl}
	\Jmongrid(k) & = & 
	\sumnNp & 
	\big(   
	\cgriddiff\nk \cdot \max(0, \Pgrid\nk) \IEEEnonumber\\ 
	&&& +\> \! \cgridsell\nk \cdot \Pgrid\nk  
	\big) \! \cdot \! \Tsamp\nk.
	\IEEEeqnarraynumspace \label{eq:Jmongrid}
\end{IEEEeqnarray}
Additionally, a penalty for the highest power peak has to be paid, as is common in German industry pricing. 
It can be expressed as \cite{schmitt2020multi,schmitt2022multi}
\begin{IEEEeqnarray}{rCl}
	\ctildepeak\nk & = &  \cgridpeak \cdot \Pgrid\nk - \Jmonpeakprev\k \IEEEeqnarraynumspace \\
	\Jmonpeak\k  & = & \max_{n=0,\ldots, \Np-1}\left( 
		\max\left(0,~ \ctildepeak\nk \right)
	\right), 
	\IEEEeqnarraynumspace \label{eq:Jmonpeak}  
\end{IEEEeqnarray}
with $\cgridpeak = 122.07\,\meuroperkwnice$. 
\Jmonpeakprev is implemented as a parameter which is calculated before each time step and denotes the already occurred peak costs.

The total monetary costs are then 
\begin{IEEEeqnarray}{rCl}
	\Jmontotal(k) & = & \Jmongrid\k + \Jmonpeak\k. 
	\label{eq:Jmontotal}
\end{IEEEeqnarray}

\subsubsection{Operational Costs}
\label{subsubsec:operational_costs}

Frequent startups increase an electrolyzer's degradation. 
Thus, we punish every startup of the electrolyzer by 
\begin{IEEEeqnarray}{rCl}
	\toggleonoff(k) & = & \belyon(k) - \belyon(k-1), \label{eq:Jstartup_2} \\
	\Jstartup(k) & = &  \sumnNp \max\big(0, ~\cstartup \cdot \toggleonoff\nk \big) 
	\IEEEeqnarraynumspace \label{eq:Jstartup} 
\end{IEEEeqnarray}
with $\cstartup = 10$. 

\subsubsection{User Satisfaction}
To enforce that any fuel demand is satisfied as much as possible, we punish the mismatch slack variable \zfuelmismatch from \eqref{eq:slack_fuel} strongly, \ie 
\begin{IEEEeqnarray}{rCl}
	\Juser(k) & = &  \cfueldemandmis \sumnNp \zfuelmismatch\nk \cdot \Tsamp\nk, \label{eq:Juser} 
\end{IEEEeqnarray}
with $\cfueldemandmis = 200\,\nicefrac{\meuro}{\kg}$. 

\subsubsection{$\CO$ Emissions}
We punish $\CO$ emissions separately, \ie 
\begin{IEEEeqnarray}{rCl}
	\Jco(k) & = &  \sumnNp \! 
	\cgridco \! \cdot \max(0,~ \Pgrid\nk) 
	\cdot \Tsamp\nk,  
	\IEEEeqnarraynumspace \label{eq:Jco} 
\end{IEEEeqnarray}
with $\cgridco = 0.02\,\meuroperkwhnice$.

\subsection{Optimal Control Problem (OCP)}
\label{sec:control_approach-ocp}

The objective function of the \gls{ocp} is the sum of all costs described above, \ie 
\begin{IEEEeqnarray}{rCl}
	\Jtotal(k) & = &  \Jsclp\k \! + \! \Jscmp\k \! + \! \Jmongrid(k) \! + \! \Jmonpeak\k \IEEEnonumber \IEEEeqnarraynumspace \\ 
	&& +\> \Jstartup(k) + \Juser(k) + \Jco(k). 
	\IEEEeqnarraynumspace \label{eq:Jtotal} 
\end{IEEEeqnarray}
Note that various approaches exist to derive appropriate weights for the individual objectives, if user preferences should be respected \cite{schmitt2022incorporating,schmitt2020application}.
The decision variables are the actual inputs, 
$\uvek = \mylinevec{\Pely, & \belyon, & \bcomplpmp, & \bcomppr, & \mdotfuelmp, & \Pgrid}$ 
as well as the auxiliary variables 
$\uaux = \mylinevec{\btildeelyon, & \zfuelmismatch, & \ldots}$ 
with all the additional variables from the \gls{pwla} formulations and logical AND conditions. 
Let \useq and \uauxseq denote the sequences of these vectors, \eg 
$\useq\k = \mylinevec{\uvek(0|k),& \ldots\,,& \uvek(\Npred-1|k)}$. 
The prediction horizon is 7 days long and split into $\Npred = 35$ unequal steps, \ie
\begin{IEEEeqnarray}{rCl}
	\Tsampseq & = & 
		\big( 
			5\minutes,~  
			10\minutes,~
			15\minutes,~
			3\,*\,30\minutes, 
			\IEEEnonumber \\
			&& 
			~~ 
			22\,*\,1\hours,~
			2\,*\,12\hours,~
			5\,*\,24\hours 
		\big),
		\IEEEnonumber
	\label{eq:pred_horizon} 
\end{IEEEeqnarray}
where $a\,*\,b$ denotes $a$ steps of length $b$. 
Then, the \gls{ocp} is given by 
\begin{IEEEeqnarray}{ccl} 
	\min_{\useq, \uauxseq} & \, & \Jtotal\k 
	\IEEEeqnarraynumspace 
	\label{eq:opt_prob_mpc_without_allocator} 
	\\
	\st & &\eqref{eq:bcomp_logic}, \eqref{eq:slack_fuel}, \eqref{eq:dyn_lp}, \eqref{eq:dyn_mp}, \eqref{eq:bal_eq} ~\forall\, n = 0 \ldots \Npred -1 \IEEEnonumber* \\ 
	&& \eqref{eq:m_lp_box}, \eqref{eq:m_mp_box},   ~\forall\, n = 1 \ldots \Npred,  \IEEEnonumber \\
	&& 	\text{constr. by 1D\,PWLAs on } \mdotely(\Pely), \mdotlpmp(\mlp), 
	\IEEEnonumber \\ 
	&& \text{constr. by 2D\,PWLA on } \Pcomplpmp(\mlp, \mmp), \IEEEnonumber \\
	&& \text{constr. by AND conditions on } \btildeelyon~\&~\belyon, \IEEEnonumber
\end{IEEEeqnarray} 
where the time step notation $(k)$ and $(k+1)$ in \eqref{eq:m_lp_box}--\eqref{eq:bal_eq} are to be read as $(n|k)$ and $(n+1|k)$, respectively. 

\subsection{Allocator}
\label{sec:control_approach-allocator}

\begin{figure}[htb]
	\centering
	\textbf{\small Simulation Flow (Including Allocator)}\par 
	\vspace{2mm}
	\includegraphics{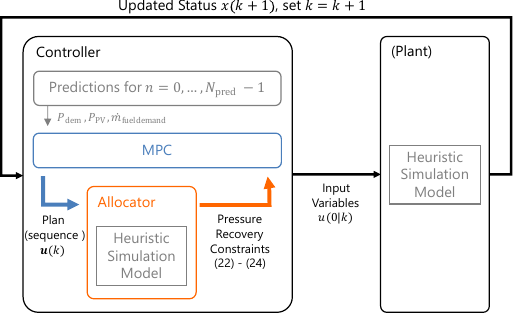}	
	\caption{
		The allocator uses a second, heuristic simulation model, which models all 6 \gls{mp} tanks individually. 
		Using the input variables determined by the \gls{mpc} solving \eqref{eq:opt_prob_mpc_without_allocator}, it checks whether all planned car refueling events would be successful, \ie if at least one tank with enough pressure is present. 
		If not, it sets the parameters for the constraints \eqref{eq:tpr}--\eqref{eq:mp_lim_alloc} and the \gls{mpc} runs a second time. 
		Afterwards, the first input variables of the updated plan are sent to the plant.
		In this work, we use the same heuristic simulation model as in the allocator as the plant. 
		Finally, the state is updated and the entire process repeated for the next time step.
	}
	\label{fig:alloc_approach}
\end{figure}

The \gls{mpc} is unaware of the individual \gls{mp} tanks' pressure levels. 
However, for the \gls{hrs}, it is crucial whether at least a single \gls{mp} tank has enough pressure to fuel a car. 
Thus, in this paragraph we present the proposed allocator approach. 
\figref{fig:alloc_approach} gives an overview of the entire simulation flow. 
\algoref{algo:heuristic_simulation_model_step} describes the differences in the iteration of the heuristic simulation model to the \gls{mpc}'s internal model. 
\algoref{algo:allocator} describes how the allocator determines whether additional constraints on the \gls{ocp} are necessary. 

\begin{algorithm}
	\caption{
		Excerpt of the heuristic simulation model's iteration. 
		Parts which are identical with the \gls{mpc}'s internal model are omitted. 
	}
	\label{algo:heuristic_simulation_model_step}
	\textbf{Inputs:} Decision variables \uvek, all tank masses, \Tsamp
	
	$\letterhighersec =$ section with higher mass, 
	$\letterlowersec =$ section with lower mass\; 
	\tcp{Do Pressure Recovery} 
	\If{$\bcomppr = 1$ and $\mseci[1] \neq \mseci[2]$} { 
		$\mprplan = \mdotprnom \cdot \Tsamp$\;
		Determine tank order (ascending masses) in lower section:
		$\letterlowersec_1, \letterlowersec_2, \letterlowersec_3$\;
		$\mpravailable =$ Available hydrogen available for PR (before either lower or higher limit of tanks/sections is hit)\; 
		$\mprtomove = \min(\mprplan, \mpravailable)$\;
		\For{$i$ in $\{\letterlowersec_1, \letterlowersec_2, \letterlowersec_3\}$} { 
			$\mprtemp = \min(\mprtomove,~ \mmpi - \mmpimin)$\;
			\tcp{Fill section $\letterhighersec$ with $\mprtemp$}
			$\mfilled = fill\_section(\letterhighersec, \mprtemp)$\;
			$\mmpi[i] \minuseq \mfilled$\;
			$\mprtomove \minuseq \mfilled$\;
			\If{\mprtemp = 0} {
				Break
			} 
		}
	}
	
	\tcp{Do HRS Refeuling}
	$\mtofuel = \mdotfuelmp \cdot \Tsamp$\;
	$\mdotfuelreal = 0$\;  
	\If{$\mtofuel > 0$} {
		Determine fuel\_order, i.e. tanks with ascending masses\;
		\For{$i$ in fuel\_order}{
			$\mdotfuelusei = \min(\mtofuel,~ \mmpi - \mmpi(p=350\,\barpressure), 0)$\;
			$\mtofuel \minuseq \mdotfuelusei$\; 
			$\mmpi \minuseq \mdotfuelusei$\; 
			$\mdotfuelreal \pluseq \mdotfuelusei/\Tsamp$\;
			\If{$\mtofuel = 0$}{
				Break		
			}
		} 
	}
	
	\tcp{Do Filling LP -> MP} 
	$\mtofill = \mdotlpmp \cdot \Tsamp$\; 
	\If{$\mtofill > 0$} {
		$\mfilledhighersec = fill\_section(\letterhighersec, \mtofill)$\;
		$\mtofill \minuseq \mfilledhighersec$\;
	}
	\If{$\mtofill > 0$} {
		$\mfilledlowersec = fill\_section(\letterlowersec, \mtofill)$\;
		$\mtofill \minuseq \mfilledlowersec$\;
	}
\end{algorithm}

\begin{algorithm}
	\caption{Allocator logic to determine additional constraints on \gls{pr} and \mmp. 
	}
	\label{algo:allocator}
	\textbf{Inputs:}  $\useq\k$ \\
	\textbf{Outputs:} $\bconsiderprseq\k$, $\tpr\k$, $\mmplimallocseq\k$
	
	Get state sequence \xseq and adjusted input sequence \useqadj from heuristic simulation model (see \algoref{algo:heuristic_simulation_model_step})\;
	Calculate (unplanned) mismatches \mdotfuelmismatchseq between \mdotfuelmpseq from \useq and \useqadj\;
	\If{$\sum_{n}{\mdotfuelmismatch(n|k)} > 0$}{
		Determine first step \nfirstmismatch in prediction horizon for which $\mdotfuelmismatch(\nfirstmismatch|k) > 0$\;
		Determine how much hydrogen \mavaforpr would be available for PR at this time\;
		\If{$\nfirstmismatch \leq \kcutoff \AND \mavaforpr(\nfirstmismatch|k) > 0$ }{
				$\bconsiderpr(n) \gets 1~\forall n = 0, \ldots, \nfirstmismatch-1$\,
				Determine \mmovewithpr as minimum of $\mavaforpr(\nfirstmismatch|k)$ and necessary amount to have higher section at $350\,\barpressure~+$ fuel demand\;
				$\tpr = \mmovewithpr / \mdotprnom$\;
				$\tpr \gets \min(\tpr, \sum_{n=0}^{\nfirstmismatch}\Tsamp(n))$\;  
			}
		Determine possible upper limits \mmppossibleseq with maximum hydrogen production\;
		$\mmpthreefiftylimit =$ lower limit to have at least one section with $350\,\barpressure$\;
		\For{every $n$ with $\mdotfuelmismatch(n|k) > 0$}{
			$\mmplimalloc(n|k) \gets \min(\mmppossible(n|k), \mmpthreefiftylimit+\mdotfuelmp(n|k) \cdot \Tsamp(n))$\;
		}
	}
\end{algorithm}

The allocator sets constraints on the \gls{pr} only in the first $8\,\hours$ of the horizon, which corresponds to the first $\kcutoff = 12$ steps,
\begin{IEEEeqnarray}{c}
	\sum_{k=0}^{\kcutoff-1} \bconsiderpr\k \cdot \bcomppr\k \cdot \Tsampk  ~ \geq ~ \tpr, \IEEEeqnarraynumspace \label{eq:tpr}
	\\
	\bconsiderpr\k  =  0    ~~\forall\, k = \kcutoff \ldots \Npred-1, \IEEEeqnarraynumspace \label{eq:mp_lim_alloc} 
\end{IEEEeqnarray}
where $\tpr \geq 0$ in hours defines how long the \gls{pr} should run. 
Additionally, a higher limit on \mmp may be set to ensure that enough hydrogen is available to be shifted between the sections, 
\begin{IEEEeqnarray}{rCl}
	\mmp\k & \geq & \mmplimalloc\k - \zlimalloc\k  \IEEEnonumber \\ 
	&& \qquad \qquad \qquad  \forall\, k = 0 \ldots \kcutoff - 1. 
	\label{eq:mp_lim_alloc} 
\end{IEEEeqnarray}
\zlimalloc is used to soften the constraint in combination with a cost function analogous to \eqref{eq:sc_lp} with a weight of $1$.
If constraints on \tpr and \bconsiderpr or on \mmplimalloc are set, the \gls{mpc} re-solves the \gls{ocp} \eqref{eq:opt_prob_mpc_without_allocator} including the new constraints.

%% file: inc/04_simulation_results.tex
\section{Simulation Study}
\label{sec:simulation_study}

To evaluate the proposed control approach, we performed a simulation study, where we simulated the controller against the heuristic model used in the allocator, as proposed in \secref{sec:control_approach-allocator}.
This is illustrated in \figref{fig:alloc_approach}.

In the first subsection, we briefly describe the simulation setup.
To benchmark the proposed approach, we compare it to two rule-based control strategies which are described in \secref{sec:simulation_study-rule_based_controllers}.
Finally, we discuss the results of the study in \secref{sec:simulation_study-results}.

\subsection{Setup}

For the simulation, real measurement data of the Honda Europe R\&D facility in Offenbach, Germany was used for the load demand and PV power.
The hydrogen demand profile for the \gls{hrs} was generated using a statistical usage scenario:
We generate a set of $\Nsessions$ random \gls{fcv} fueling sessions for each calendar week of the year, where $\Nsessions \sim \mathcal{U}[5, 10]$.
\glspl{fcv} arrive with equal probability on Monday -- Friday, while no cars arrive on weekends.
For each session, we draw an arrival time \tarr and a hydrogen demand \mhtwo.
50\percent of the sessions occur between 7~--~9:30 a.m., with 10\percent between 12~--~1 p.m., and 40\percent between 4:30~--~6:30 p.m.
Therefore, a time window for $\tarr$ is selected accordingly.
Within the selected window, $\tarr$ is distributed uniformly.
The demand is sampled as $\mhtwo = \max(4\,\kg, \mhtwotilde)$, with $\mhtwotilde \sim \mathcal{N}(3\,\kg, (0.5\,\kg)^2)$.
Each fueling session is assumed to take $5\minutes$ and the fuel demand rate $\mdotfueldemand$ is calculated accordingly.
In total, 410 fueling sessions were generated for the simulated year.
The process of generating sessions was adapted from \cite{engel2022hierarchical}.
	
As initial states of the tanks, we assume $\mlp\nul = 5\,\kg$,
$\mmpi[1]\nul = 83\,\kg$,
$\mmpi[2]\nul = 82\,\kg$,
$\mmpi[3]\nul = 81\,\kg$,
$\mmpi[4]\nul = 80\,\kg$,
$\mmpi[5]\nul = 60\,\kg$,
$\mmpi[6]\nul = 60\,\kg$.
Furthermore, the initial peak limit is $\Pgridpeak\nul = 500\,\kW$ and the electrolyzer is off, \ie  $\belyon\nul = 0$.

We simulated the full calendar year of 2021 using perfect predictions.
An analysis of the influence of prediction errors for the \gls{pv} power output on the peak costs can be found in \cite{schmitt2021cost}. 
The controller, model and simulation framework were implemented in Python, using CVXPY \cite{diamond2016cvxpy} to model the \gls{ocp} and to interface Gurobi \cite{gurobi} as the solver.
The simulation step size is $\Tsamp(0|k) = 5\minutes$.
We simulated on an Intel i5-9400, with a solver time limit of 20\seconds and an optimality gap of $10^{-4}$.
The MPC simulation took approx. $11\hours$, and the rule-based controller simulations approx. $12\minutes$ each.

\subsection{Rule-Based Benchmark Controllers}
\label{sec:simulation_study-rule_based_controllers}
\glsunset{rbcp} 
\glsunset{rbce}

To benchmark the performance of the proposed approach, we propose two baseline rule-based controllers.
The first controller (\gls{rbcp}) uses as much power as is available without causing a new peak to operate the electrolyzer, \ie
\begin{IEEEeqnarray}{rCl}
	\Pavailablepeak\k &=& \max \big(\Pgridpeakk + \Ppv\k \IEEEnonumber\\ 
	&& \quad \quad \quad +\> \Pdemk - \Pcompmax, 0\big).
\end{IEEEeqnarray}
The second controller (\gls{rbce}) uses only available excess PV power, \ie
\begin{IEEEeqnarray}{rCl}
	\Pavailableexcess\k &=& \max \big(0, \Ppv\k + \Pdemk\big).
\end{IEEEeqnarray}
Both controllers employ the same logic to apply $\Pavailable(k)$ to operate the plant.
This logic is illustrated in \algoref{algo:rbc_logic}.
Note that the \gls{rbce} controller may also use non-PV excess power to run the compressor or perform pressure recovery.
As for the MPC, $\Tsamp = 5\,\minutes$ is used for both controllers.

\begin{algorithm}
	\caption{Operational logic of rule-based controllers}
	\label{algo:rbc_logic}
	
	$\melymin\k = \mdotely(\Pelymin) \cdot \Tsamp$\;
	$\melymax\k = \mlpmax - \mlp\k$\;
	
	\If{$\Pavailable(k) \geq \Pelymin \AND \melymax\k \geq \melymin\k$}{
		$\belyon\k \gets 1$\;
		
		\eIf{$\belyready\k = 1$}{
			$\Pmax\k = \max \left(\Pely(\melymax\k / \Tsamp), \Pelymax \right)$\;
			$\Pely\k \gets \min \big(\Pavailable\k, \Pmax\k\big)$\;
		}{
			 $\Pely\k \gets 0$
		}
	}
	
	$\Pres\k = \Pavailable\k - \Pely\k$\;
	\If{$\Pdemk - \Pcompmax + \Pres\k \leq \Pgridpeakk$}{
		\eIf{Transfer from $\g{LP} \rightarrow \g{MP}$ possible}{
			$\bcomplpmp\k \gets 1$
		}{
			$\bcomppr\k \gets 1$
		}
	}
	
	$\mdotfuelmp\k \leftarrow \mdotfueldemand\k$
\end{algorithm}

\subsection{Simulation Results}
\label{sec:simulation_study-results}

\tabref{tab:kpis} shows the results of the simulation study in terms of yearly \glspl{kpi}.
The electricity costs do not contain the peak costs.
The PV self consumption rate is defined as the fraction of PV energy that is not fed back into the grid. 
The costs per $\kg$ of hydrogen are calculated from the electricity costs incurred by the operation of any of the hydrogen components (\ie electrolyzer, compressor).
Here, the energy that could be covered with PV excess is accounted for with the feed-in tariff $\cgridsell$, and power from grid with the grid tariff $\cgridbuy$.
The fueling success rate is the fraction of fulfilled hydrogen demand and total hydrogen demand.

\begin{figure}[htb]
	\centering
	\includegraphics{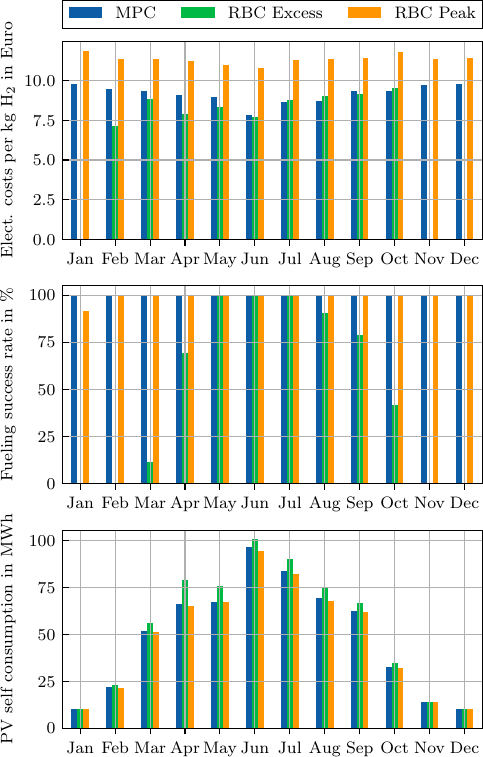}	
	\caption{Electricity costs per $\kg$ hydrogen, fueling success rate and PV self consumption in each month of the simulated year 2021 for the proposed MPC approach, the RBC Excess controller and the RBC Peak controller.}
	\label{fig:kpis_over_year}
\end{figure}

The results in \tabref{tab:kpis} show that the \gls{rbce} controller incurs the lowest electricity costs at the highest PV self consumption rate.
As the controller uses mostly excess PV power, it naturally incurs the lowest costs of hydrogen production and lowest $\g{CO}_2$ emissions.
Yet, this comes at the cost of the fueling success rate being very low.
This is due to the total amount of hydrogen produced being half of what the \gls{rbcp} or MPC controllers produce, indicating that for the given plant, excess PV power is not sufficient to fulfill hydrogen demand.
This is further illustrated in \figref{fig:kpis_over_year}.
The figure shows the costs per $\kg$ hydrogen, fuel success rate and PV self consumption in each month of the simulated year.
In the winter months, the comparatively low amount of available PV energy is not enough to fulfill the hydrogen demand, indicated by the very low fuel success rate of the \gls{rbce} controller.

Compared to the \gls{rbcp} controller, the proposed MPC approach incurs lower electricity costs while also producing hydrogen at a significantly lower cost.
The hydrogen production costs of the MPC are only slightly higher than those of the \gls{rbce} controller, while achieving a significantly higher fuel success rate of 100\percent.
The \gls{rbcp} controller produces approx. $90\,\kg$ hydrogen more than the MPC at a lower overall PV self consumption rate.
Furthermore, the \gls{rbcp} controller produces 49 more startups of the electrolyzer compared to the MPC.
While the average number of startups per day is still less than 1, the total number of startups should be minimized, as these degrade the hydrogen stacks.

The \gls{rbcp} controller exhibits a decreased fuel success rate in January, as can be seen in the middle plot of \figref{fig:kpis_over_year}.
This is due to the initial tank conditions.
The rule-based controllers only perform pressure recovery if no electrolyzer operation is possible and/or no peak power demand would be caused.
In January, this behavior in combination with the initial conditions causes the \gls{rbcp} controller to not perform pressure recovery in the beginning of the month, whereby the demand of the first two cars arriving cannot be fulfilled completely.

\begin{table}[h]
	\caption{\glspl{kpi} of the proposed control approaches for the year 2021.}
	\label{tab:kpis}
	\begin{tabular}{lrrr}
		\toprule
		{} &     MPC &  RBC Excess &  RBC Peak \\
		KPI                                   &         &             &           \\
		\midrule
		Electricity costs in 1000 Euro        &  249.43 &      244.45 &    253.25 \\
		Max. peak in kW                       &  544.88 &      544.88 &    544.88 \\
		$\mathrm{CO}_2$ emissions in kg       &  760.57 &      735.87 &    773.58 \\
		Produced $\mathrm{H}_2$ in kg         & 1245.77 &      752.36 &   1336.01 \\
		Elect. costs per kg $\mathrm{H}_2$ in Euro   &    9.17 &        8.54 &     11.41 \\
		Fueling success rate in \%            &  100.00 &       54.49 &     99.58 \\
		PV self consumption in \%             &   86.42 &       93.41 &     85.24 \\
		PV self consumption in $\mathrm{MWh}$ &  587.38 &      634.85 &    579.34 \\
		Number of electrolyzer startups       &  297 &      289 &    348 \\
		\bottomrule
	\end{tabular}

\end{table}

%% file: inc/05_conclusion.tex
\section{Conclusion}
\label{sec:conclusion}
We proposed an \gls{mpc} approach with implicit incorporation of heuristics for the operation of a hydrogen plant.
From the overall results of the simulation study, we conclude that the proposed approach operates the plant effectively and efficiently.
Furthermore, we have shown that it outperforms the two proposed rule-based benchmark controllers. 

While the simulation study showcases the efficacy of the proposed MPC approach, this should be validated on the actual hardware of the hydrogen plant.
As an intermediate step, the approach could be validated through \gls{sil} simulation using a high-fidelity digital twin of the facility, as was done in \cite{schmitt2023errorcomp}.
Thus, future work will involve the implementation and testing of the proposed MPC controller in \gls{sil} simulation on a digital twin and on the physical plant.